\documentclass[conference]{IEEEtran}
\IEEEoverridecommandlockouts
\usepackage{caption}
\usepackage{algorithm}
\usepackage{algorithmic}
\usepackage{graphicx}
\usepackage{enumerate}
\usepackage{amsthm}
\usepackage{epstopdf}
\usepackage{amsmath,amssymb,epsfig,graphicx,slidesec,colortbl}
\usepackage[absolute,overlay]{textpos}
\usepackage{multimedia,algorithm,algorithmic,multirow,url, bbm}
\usepackage{cite}
\usepackage{subfig}
\usepackage{color}
\usepackage{url}
\usepackage{enumitem}

\newtheorem{Prob}{Problem}

\newtheorem{The}{Theorem}
\newtheorem{Pro}{Proposition}

\newtheorem{Assump}{Assumption}
\newtheorem{Lem}{Lemma}
\newtheorem{Obs}{Observation}
\newtheorem{Que}{Question}

\usepackage{geometry}
\title{Incentive Mechanism Design for Distributed Ensemble Learning}
\author{Chao Huang*, Pengchao Han*, and Jianwei Huang, \textit{IEEE Fellow}
\thanks{*Equal Contribution.}
\thanks{Chao Huang is with the Department of Computer Science, the University of California,
        Davis, USA. Email:fchhuang@ucdavis.edu. Pengchao Han is with the School of Science and Engineering, The Chinese University of Hong Kong, Shenzhen, China. Email:hanpengchao@cuhk.edu.cn. Jianwei Huang (corresponding author) is with the School of Science and Engineering, The Chinese University of Hong Kong, Shenzhen, and the Shenzhen Institute of Artificial Intelligence and Robotics for Society. Email: jianweihuang@cuhk.edu.cn.}
\thanks{This work is supported by the National Natural Science Foundation of China (Project 62271434), Shenzhen Science and Technology Program (Project JCYJ20210324120011032), Guangdong Basic and Applied Basic Research Foundation (Project 2021B1515120008), Shenzhen Key Lab of Crowd Intelligence Empowered Low-Carbon Energy Network (No. ZDSYS20220606100601002), and the Shenzhen Institute of Artificial Intelligence and Robotics for Society. The work of Pengchao Han is supported by Guangdong Basic and Applied Basic Research Foundation under Grants 2022A1515110056.}
}
\begin{document}
\pagenumbering{gobble}
\maketitle

\begin{abstract}
 Distributed ensemble learning (DEL) involves training multiple models at distributed learners, and then combining their predictions to improve performance. Existing related studies focus on DEL algorithm design and optimization but ignore the important issue of incentives, without which self-interested learners may be unwilling to participate in DEL. We aim to fill this gap by presenting a first study on the incentive mechanism design for DEL. Our proposed mechanism specifies both the amount of training data and reward for learners with heterogeneous computation and communication costs. One design challenge is to have an accurate understanding regarding how learners' diversity (in terms of training data) affects the ensemble accuracy. 
 To this end, we decompose the ensemble accuracy into a diversity-precision tradeoff to guide the mechanism design. Another challenge is that the mechanism design involves solving a mixed-integer program with a large search space. To this end, we propose an alternating algorithm that iteratively updates each learner's training data size and reward. We prove that under mild conditions, the algorithm converges.  Numerical results using MNIST dataset show an interesting result: our proposed mechanism may prefer a lower level of learner diversity to achieve a higher ensemble accuracy.
\end{abstract}

\section{Introduction}

The wisdom of the crowd refers to the often observed phenomenon that the collective knowledge of a group of individuals is often more accurate than that of an  expert. Ensemble learning is a machine learning interpretation of such a phenomenon that involves combining multiple learning models to improve the overall predictive performance and robustness.  Ensemble learning methods, such as bagging, boosting, and stacking, have been successfully applied in various sectors, including finance, healthcare, and transportation \cite{sagi2018ensemble}. 

\begin{figure}[t]
	\centering
 \includegraphics[width=0.96\linewidth]{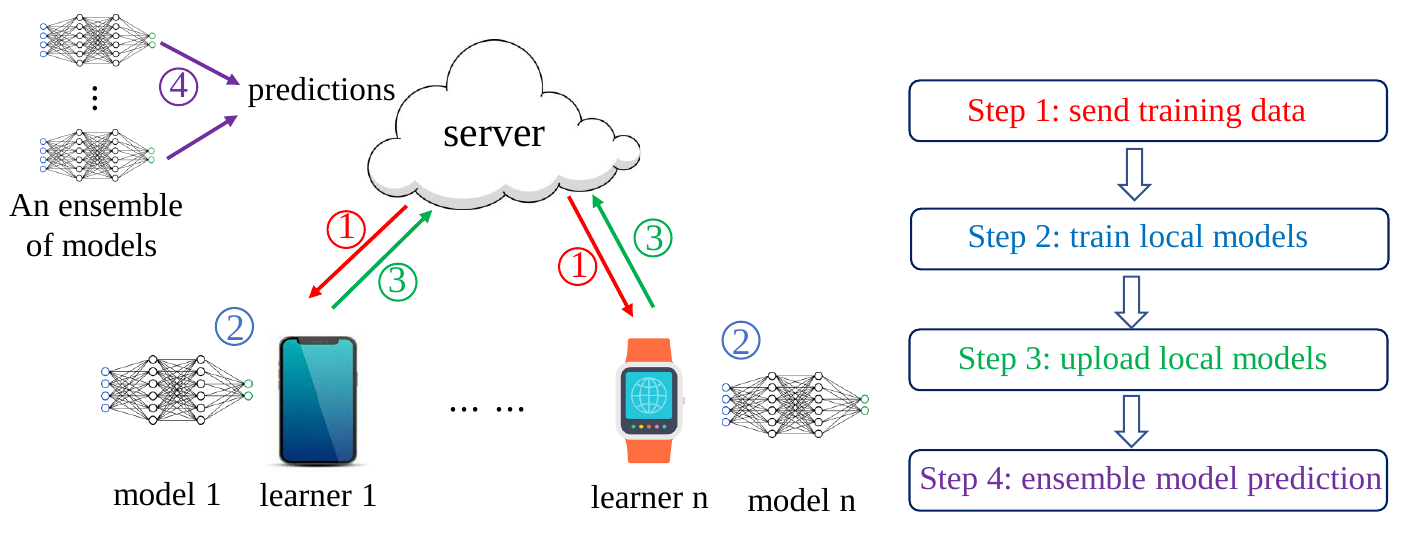}	
	\caption{Distributed ensemble learning (e.g., bagging).}
	\label{DEL}
\end{figure}

Despite its improved performance and robustness, ensemble learning can be computationally intensive, as it involves training multiple models and then combining their predictions \cite{qiu2019ensemble}. The overall computational burden increases with the number of models, the size of the training data, and the complexity of the models. This can be a significant challenge, particularly when dealing with large datasets or complex models. 
A promising solution is distributed ensemble learning (DEL), in which a central server coordinates the training of an ensemble of models across multiple distributed learners (e.g., IoT devices, mobile phones, and edge servers) \cite{tekin2016adaptive}. A typical DEL process consists of four steps (see also Fig. \ref{DEL}):
\begin{itemize}
\item \textbf{Step 1}: The server samples subsets of data from a large dataset and sends them to respective learners. 
\item \textbf{Step 2}: The learners train machine learning models in parallel using their downloaded datasets.
\item \textbf{Step 3}: The learners upload trained models to the server.
\item \textbf{Step 4}: The server combines the models into an ensemble model and uses it to produce final predictions.
\end{itemize}
In DEL, learners can train on  smaller subsets of data in parallel, leading to faster overall training time. 


There has been some excellent work on the algorithmic design of DEL. One area of focus is developing more efficient and scalable distributed learning frameworks (e.g., parameter servers and data/model parallelism) that can improve the training time and resource utilization \cite{ding2016towards,qin2021solving}. Another area of focus is improving the robustness and generalization capabilities by developing model selection/pruning methods \cite{bian2021subarchitecture,bian2019ensemble}. However, these prior studies ignored the important issue of incentive design. Specifically, training at the distributed entities requires costly computation (and communication for data/model transmission). Without proper incentives, the entities may not be willing to participate and faithfully perform model training. This paper takes a first attempt to answer the question below: 
\begin{Que}
How to design an effective incentive mechanism for distributed ensemble learning? 
\end{Que}

To answer Question 1, we consider a scenario where a central server aims to incentivize distributed learners to participate and finish model training tasks. The server aims to \emph{maximize a tradeoff between the ensemble model accuracy and the total costs of incentivizing learners}. 
The incentive design for DEL is challenging due to two reasons as follows.

First, \textit{diversity} is the key to achieving a good ensemble model accuracy \cite{zhang2022novel}. That is, 
individual models should be diverse and complement each other's strengths and weaknesses, leading to more accurate and robust predictions. However, there is still no consensus till today in the research community on how to best measure diversity and how diversity affects the ensemble model accuracy \cite{bian2021does}. To address this issue, motivated by \cite{bian2021does}, we proceed from a diversity-precision decomposition perspective and define a surrogate function to simulate the true ensemble accuracy. The surrogate function contains two parts: (1) ``diversity'' that is measured by the number of  mistakes that learners make during prediction; (2) ``precision'' that reflects the average performance of learners on their own datasets. The use of the surrogate function presents an important tradeoff between learners' diversity and precision, which will be helpful in guiding the incentive design.

Second, distributed learners usually have heterogeneous computation costs (for model training) and communication costs (for data downloading and model uploading). This requires a customized design of the learning task (i.e., training data) and the reward for each learner, resulting in a mixed-integer program with a huge search space.
To address this issue, we propose an alternating algorithm that updates the reward and the training data for each learner in a round-robin fashion. 
As will be shown, our proposed algorithm significantly reduces the search space and achieves fast convergence.   
 







\subsection{Key Contributions}
The key contributions of this paper are as follows.
\begin{itemize}
\item \textit{Incentive design for distributed ensemble learning}: To our best knowledge, this is the first attempt to study the incentive mechanism design for distributed ensemble learning. We propose an incentive  mechanism that specifies both amount of training data and reward for learners with heterogeneous computation and communication costs.  


\item \textit{Alternating optimization algorithm}: The incentive design involves solving a challenging mixed-integer problem with a huge search space. To this end, we propose an alternating algorithm that updates each learner's reward and training data in a round-robin fashion. The algorithm greatly reduces the search space and is provable convergent. It also has a polynomial complexity in terms of the number of learners, and hence is scalable to large distributed systems.

\item \textit{Numerical experiments:} We conduct experiments using MNIST \cite{mnist}. Our results also reveals an interesting interaction between learner diversity and the ensemble accuracy. Specifically, the mechanism may prefer a lower level of learner diversity to achieve a higher ensemble accuracy. 
\end{itemize}

The remainder of this paper is organized as follows. We present the system model in Section \ref{sec-model}. We provide theoretical analysis in Section \ref{sec-analysis}. We present numerical experiments in Section \ref{numerical} and conclude in Section \ref{sec: conclusion}. 
\section{System Model}\label{sec-model}

We first present the system model for the distributed learners' decision problem in Section \ref{learner-decision-pro}, and then turn to the server's mechanism design problem in Section \ref{server-decision}.
\subsection{Learners' Decision Problem}\label{learner-decision-pro}
In this subsection, we first introduce the task and  learners. Then, we define each learner's strategy and payoff function, and formulate its decision problem. 

\subsubsection{Learners and Task}
There is a set $\mathcal{N}=\{1, 2, \cdots, N\}$ of learners (e.g., mobile devices) that can be reached by the server. The task of each learner $i\in \mathcal{N}$ is to train a classification model using data provided by the server. Define:
\begin{itemize}
\item $\mathcal{M}_i$: learner $i$'s machine learning model (e.g., multi-layer perceptron) with a model size  $M_i =|\mathcal{M}_i|$.
\item $\mathcal{D}_i$: learner $i$'s training data that is chosen by the server,
with the data size $D_i=|\mathcal{D}_i|$.
\end{itemize}
After the local training process converges, each learner $i$ sends the trained model $\mathcal{M}_i$ to the server for downstream analysis.  

\subsubsection{Learner Participation Strategy}
Each learner $i$ decides whether to participate in distributed ensemble learning to perform the training task. We use a binary variable $d_i\in \{0,1\}$ to denote a learner's participation decision, where $d_i=1$ means participating and $d_i=0$ means not participating.\footnote{In this paper, we assume that if a learner $i$ participates, it will faithfully perform the training task using data $\mathcal{D}_i $ and truthfully upload the trained model $\mathcal{M}_i$. This is reasonable, as the server can verify the performance of learners' uploaded models using a held-out dataset.} 

\subsubsection{Computation and Communication Costs}
A participating learner mainly incurs two types of costs: computation cost and communication cost, which we elaborate on as follows.

\textit{Computation cost}: Performing model training  consumes computation
resources. Let $C_i^{\rm comp}$ denote the computation cost, which is a linear function of learner $i$'s data size \cite{zhang2022enabling}:
\begin{equation}\label{cost-computation}
C_i^{\rm comp}=\alpha_i D_i.
\end{equation}
The computation cost coefficient of learner $i$, $\alpha_i>0$, depends on various factors such as the learner's computing chip architecture and CPU processing speed. 

 \textit{Communication cost}: A learner needs to consume communication resources (e.g., using wireless networks) for  downloading training data from and  uploading trained model to the server. Let $C_i^{\rm comm}$  denote learner $i$'s communication cost:
\begin{equation}
C_i^{\rm comm}= \beta_i (D_i + M_i),
\end{equation}
where $\beta_i>0$ represents learner $i$'s communication cost coefficient that depends on the channel conditions. For convenience, we normalized $M_i$ to zero, as it is much smaller than $D_i$ in many settings. For example, in our experiments on MNIST dataset, the training data size $D_i$ is 46.4M, while the neural network model size $M_i$ is only 1.7M. One can easily extend our analysis to the case  where $|\mathcal{M}_i|$ is non-negligible. 


\subsubsection{Reward}
Without enough incentives, learners may not be willing to participate in DEL. The server provides a reward $R_i\ge0$ to each participating learner $i$ to compensate the computation and communication costs. For non-participating learners, the server does not provide any reward.

\subsubsection{Learner Payoff Maximization Problem} We define each learner $i$'s payoff function as:
\begin{equation}\label{learner-payoff}
\begin{aligned}
u_i(d_i; R_i, D_i) = \begin{cases}
R_i - \alpha_i D_i - \beta_iD_i, \quad &\text{if} \quad d_i=1,\\
0, \quad &\text{if} \quad d_i=0.
\end{cases}
\end{aligned}
\end{equation}
Given $R_i$ and $D_i$, each learner $i$ decides $d_i$ to maximize its payoff. The problem is formulated below.
\begin{Prob}{(Leaner $i$'s Participation Problem)}
\begin{equation}\label{learner-problem}
\begin{aligned}
\max & \quad u_i(d_i; R_i, D_i) \\
     {\rm var}. &\quad d_i\in \{0,1\}.
\end{aligned}
\end{equation}

\end{Prob}

\subsection{Server's Mechanism Design Problem}\label{server-decision}

In this subsection, we model how the server optimizes the mechanism choices for each learner to maximize its payoff, i.e., a tradeoff between the ensemble model accuracy and the total costs of incentivizing learners. 

\subsubsection{Server Mechanism Choices}
For each learner $i$, the server needs to decide the reward $R_i\ge 0$ to compensate the cost. The server also needs to decide the training dataset $\mathcal{D}_i$ for each learner. As the first attempt to study the incentive design for DEL, we focus on the widely adopted bagging (i.e., \underline{b}ootstrapped \underline{agg}regat\underline{ing}) approach \cite{whitaker2022prune}. In bagging, learners train models in parallel using bootstrapped data (sampled with replacement from the server's dataset), and the server adopts majority voting to aggregate the prediction results from all learners.\footnote{The incentive mechanism design for other ensemble approaches such as boosting and stacking will require a very different approach  and is out of the scope of this paper (e.g., in boosting, learners train models sequentially and a learner's dataset is affected by the prediction results from the previous learner).} 

With bagging, the server's decision on dataset $\mathcal{D}_i$ reduces to the datasize $D_i\in \{0, 1, 2, \cdots, D^{\rm max}\}$, where $D^{\rm max}>0$ is the size of server's available dataset. For notational convenience, we define
 $\boldsymbol{R}=\{R_i\}_{i\in \mathcal{N}}, \boldsymbol{D}=\{D_i\}_{i\in \mathcal{N}}$,
 $\boldsymbol{R}_{-i}=\{R_j\}_{j\in \mathcal{N}\setminus \{i\}}$, and $\boldsymbol{D}_{-i}=\{D_j\}_{j\in \mathcal{N}\setminus\{i\}}$.




\subsubsection{Ensemble model accuracy}
The major target of the server is to obtain an ensemble of models with good performance, i.e., the aggregated prediction results are accurate.
The key is to ensure that learners are ``diverse'' so that multiple models can complement each other's weaknesses and make fewer mistakes. However, it is difficult to analyze how the ensemble accuracy depends on learners' diversity due to several reasons:
\begin{itemize}
\item First, there is still no consensus till today in the community on how to best measure diversity \cite{bian2021does}, and how diversity affects the ensemble model accuracy. 
\item Second, learners are both heterogeneous (due to having different training data) and \emph{dependent (due to having overlapping datasets from bagging) in model precision.}   This makes a closed-form characterization of the ensemble accuracy difficult.
\end{itemize}



To address this challenge, we define a surrogate function from a diversity-precision decomposition perspective to simulate the true ensemble accuracy. We first provide some notations for ease of presentation:
\begin{itemize}
  \item $D^T(\boldsymbol{D})=|\cup_{i\in \mathcal{N}} \mathcal{D}_i|$: the size of the union of all learners' training datasets.
  \item $\mathcal{N}^P(\boldsymbol{R}, \boldsymbol{D})$: the set of participating learners, and the number of participating learners is $N^P=|\mathcal{N}^P|$.
  \item $\bar{p}(\boldsymbol{R}, \boldsymbol{D})=\sum_{i \in \mathcal{N}^P}p_i(D_i)/N^P$: learners' average precision, where $p_i$ is learner $i$' precision.
  \item $l_d(\boldsymbol{R}, \boldsymbol{D})$: the number of learners that give wrong predictions on data sample $x_d\in \cup_{i\in \mathcal{N}} \mathcal{D}_i$.
\end{itemize}

Motivated by the double fault measure in \cite{bian2021does}, we define the surrogate ensemble accuracy function:

\begin{equation}\label{transformed-double-fault}
\begin{aligned}
\tilde{F}(\boldsymbol{R}, \boldsymbol{D})=
\underbrace{\frac{\sum_{d=1}^{|\cup_{i\in \mathcal{N}} \mathcal{D}_i|}{l_d}^2}{D^T\cdot N^P\cdot(N^P-1)}}_{\rm diversity} +\underbrace{\frac{\bar{p}-1}{N^P-1}}_{\rm precision}.
\end{aligned}
\end{equation}
The first term in (\ref{transformed-double-fault}) measures the diversity. Intuitively, the learners are more diverse if they make more mistakes (e.g.,  a larger $l_d$ which likely leads to more decision boundaries). The second term reflects the average precision of learners. One can see that (\ref{transformed-double-fault}) presents an intrinsic tradeoff between diversity and precision. If learners make more mistakes, the diversity level increases but the average precision decreases. 

In what follows we will use $\tilde{F}$ as a surrogate function for the true ensemble accuracy. 
As mentioned,   $\tilde{F}$ represents a concise view of diversity-precision tradeoff that can better guide the mechanism design. Our experiments in Section \ref{DF-experiments} show that $\tilde{F}$ is indeed a good surrogate to the true ensemble accuracy. Nonetheless, one can easily extend our incentive mechanism to other surrogate functions. 

\subsubsection{Server Cost} The server's cost is the total amount of rewards allocated to learners, i.e., $\sum_{i\in \mathcal{N}} R_i$.

\subsubsection{Server Mechanism Design Problem} 
The server's payoff function is defined as the difference between the surrogate ensemble accuracy and the server's cost to incentivize learners:
\begin{equation}\label{server-payoff}
	\Pi(\boldsymbol{R}, \boldsymbol{D}) =  \gamma \cdot \tilde{F}(\boldsymbol{R}, \boldsymbol{D}) - \sum_{i\in \mathcal{N}} R_i,
\end{equation}
where $\gamma\ge 0$ represents the weight of the ensemble  accuracy. The server chooses the reward vector $\boldsymbol{R}$ and data size vector $\boldsymbol{D}$ to maximize its payoff. The problem is formulated as follows.
\begin{Prob}{(Server's Mechanism Design Problem)}\label{server-mechanism-problem}
\begin{equation}\label{server-problem}
\begin{aligned}
    \max & \quad \Pi(\boldsymbol{R}, \boldsymbol{D})\\
    {\rm var} & \quad R_i\ge 0, D_i\in \{0,1,\cdots, D^{\rm max}\}, \forall i \in \mathcal{N}.
\end{aligned}
\end{equation}
\end{Prob}

	
\section{Theoretical Analysis}\label{sec-analysis}
We first analyze each learner's optimal participation decision in Section \ref{learner-decision}. Then, we discuss how to optimize the server's mechanism design in Section \ref{server-mechanism-design}.

\subsection{Learner's Optimal Participation}\label{learner-decision}
We solve the learners' participation problem in (\ref{learner-problem}) and present the result in Lemma \ref{learner-optimal-decision}.
\begin{Lem}\label{learner-optimal-decision}
Given $\boldsymbol{R}$ and $\boldsymbol{D}$, a learner $i$'s optimal participation decision is 
\begin{equation}
\begin{aligned}
d_i^*(\boldsymbol{R}, \boldsymbol{D})=\begin{cases}
1, \quad &{\rm if} \quad R_i \ge (\alpha_i+\beta_i)D_i, \\
0, \quad &{\rm if} \quad R_i < (\alpha_i+\beta_i)D_i.
\end{cases}
\end{aligned}
\end{equation}
\end{Lem}

\textbf{Due to space limits, we only outline the sketches and defer the detailed proofs to the online appendix \cite{appendix}}. 

We prove Lemma \ref{learner-optimal-decision} by comparing the learner payoff (in (\ref{learner-payoff})) at different values of $d_i$. Lemma \ref{learner-optimal-decision} shows that a learner will participate in DEL if the provided reward $R_i$ is relatively large or the size of the training dataset $D_i$ is relatively small. 

\subsection{Server's Optimal Mechanism Design}\label{server-mechanism-design}
We achieve the server's mechanism design in three steps. First, given the data size, we optimize the reward design in subsection \ref{server-reward-design}. Then, given the reward, we optimize the data size design in subsection \ref{server-data-size-design}. Next, we discuss the joint optimization of the reward and data size in subsection \ref{server-joint-design}.

\subsubsection{Server Reward Design}\label{server-reward-design}
We summarize the server's reward design for each learner in Proposition \ref{optimal-reward}.
\begin{Pro}\label{optimal-reward}
 Given  $\boldsymbol{R}_{-i}$ and $\boldsymbol{D}$, 
the optimal reward for learner $i$ is 
	 \begin{equation}\label{reward-amount}
	 	\begin{aligned}
	 		R_i^*(\boldsymbol{R}_{-i},\boldsymbol{D}) = \begin{cases}
	 			(\alpha_i+\beta_i)D_i, \quad &{\rm if} \quad (\ref{reward-condition})\quad  {\rm holds}, \\
	 			0, \quad &{\rm else}.
	 		\end{cases}
	 	\end{aligned}
	 \end{equation}
  \begin{equation}\label{reward-condition}
  \gamma\left(\tilde{F}|_{R_i=(\alpha_i+\beta_i)D_i}-\tilde{F}|_{R_i=0}\right)\ge (\alpha_i+\beta_i)D_i.
  \end{equation}
\end{Pro}
We prove Proposition \ref{optimal-reward}  by calculating whether the benefit of learner $i$'s participation outweighs the server's cost to incentivize the learner. Proposition \ref{optimal-reward} has three implications:
\begin{itemize}
\item Proposition \ref{optimal-reward} reduces the decision space of $R_i$ from $[0, \infty)$ to binary space $\{0, (\alpha_i+\beta_i)D_i\}$. 
\item If a learner is assigned a larger dataset, or it has a larger cost coefficient, the server needs to provide a larger reward to incentivize participation (see (\ref{reward-amount})).
\item  If the server cares more about the ensemble model accuracy (i.e., a larger $\gamma$), it is more likely to incentivize learner $i$'s participation (see (\ref{reward-condition})).
\end{itemize}


\subsubsection{Server Data Size Design}\label{server-data-size-design}
Given $\boldsymbol{R}$ and $\boldsymbol{D}_{-i}$, the server solves the following problem to find the optimal $D_i$:
\begin{Prob}{(Data Size Design for Learner $i$)}\label{original-problem}
\begin{equation}
\begin{aligned}
 \max &\quad \gamma \tilde{F}(D_i) - (\alpha_i+\beta_i)D_i\\
 {\rm var} & \quad D_i\in \{0, 1,2, \cdots, D^{\rm max}\} 
 \end{aligned}
 \end{equation}
\end{Prob}

It is difficult to provide a closed-form characterization of learner $i$'s optimal data size due to it being a discrete variable. To obtain cleaner insights, we solve a relaxed continuous version of the data size design for learner $i$. More specifically, given $\boldsymbol{R}$ and $\boldsymbol{D}_{-i}$, the server solves the following problem:
\begin{Prob}{(Relaxed Data Size Design for Learner $i$)}\label{relaxed-problem}
\begin{equation}
\begin{aligned}
 \max &\quad \gamma \tilde{F}(D_i) - (\alpha_i+\beta_i)D_i\\
 {\rm var} & \quad D_i\in \left[0, D^{\rm max}\right] 
 \end{aligned}
 \end{equation}
\end{Prob}
If the optimal solution to Problem \ref{relaxed-problem} is feasible to Problem \ref{original-problem}, then it is also the optimal solution to Problem \ref{original-problem}. Otherwise, one can round the solution as an approximation. Also, the optimal objective value of Problem \ref{relaxed-problem} provides an upper bound of the optimal objective value of Problem \ref{original-problem}.

Next, we characterize some useful properties of the solutions to Problem \ref{relaxed-problem}. We start with a minor assumption.
\begin{Assump}\label{monotone-assumption}
$\tilde{F}$ is non-decreasing in $D_i$ for each $i$. 
\end{Assump}
Assumption \ref{monotone-assumption} means that the ensemble accuracy increases in a learner's data size. Our experiments in Section \ref{numerical} (e.g., Fig. \ref{tildeF}) are consistent with this assumption.

\begin{Pro}\label{monotonicity}
Under Assumption \ref{monotone-assumption}, 
(i) $D_i^*$ is  non-decreasing  in $\gamma$.
(ii) $D_i^*$ is  non-increasing in both $\alpha_i$ and $\beta_i$.
\end{Pro}
Proposition \ref{monotonicity} is proven by showing that $\partial^2 \Pi /(\partial D_i\partial \gamma) \ge 0$, $\partial^2 \Pi /(\partial D_i\partial \alpha_i) \le 0$, and $\partial^2 \Pi /(\partial D_i\partial \beta_i )\le 0$.
Proposition \ref{monotonicity} implies that if the server attaches more importance to the ensemble accuracy, it will assign a larger dataset to a learner $i$. However, it will assign less data if learner $i$ has a larger communication/computation cost coefficient. 


\subsubsection{Server Mechanism Design}\label{server-joint-design}
So far we have characterized the reward and data size design for each learner $i$, given that the design for other learners (i.e., $\boldsymbol{R}_{-i}$ and $\boldsymbol{D}_{-i}$) is fixed. These results provide guidance into the joint optimization of $\boldsymbol{R}$ and $\boldsymbol{D}$ for all learners (see Problem \ref{server-mechanism-problem}).

Next, we present an alternating optimization algorithm  that iteratively updates the reward and the data size design, as shown in Algorithm 1. Let $t\in \mathcal{Z}_+$ denote the iteration index, and the server starts with a randomized choice of $\boldsymbol{R}$ and $\boldsymbol{D}$. The server first sorts the learners based on their cost coefficients,\footnote{This corresponds to the case where the server has learners' information and can model the scenario where server and learners had previous interactions. We leave the case where such information is unknown to future work.}  and then optimizes each learner's data size (via solving Problem \ref{relaxed-problem}) and reward (via (\ref{reward-amount})-(\ref{reward-condition})) in a round-robin fashion. The algorithm terminates when the 
relative difference of the variables between consecutive interations is small. 

\begin{algorithm}[t]
	\caption{Alternating Reward and Data Size Optimization}  
	\label{alg:B}  
	\begin{algorithmic}[1] 
		\STATE {\textbf{initialization}:  let the iteration index be $t=0$. Randomly initialize $\boldsymbol{R}(t=0)$ and $\boldsymbol{D}(t=0)$.
            \STATE {\textbf{sorting}}: sort learners in  ascending order w.r.t. $\alpha_i+\beta_i$, based on which re-index learners $k = 1, 2, 3, \cdots, N$.
	\REPEAT
	\FOR{$k=1, 2,3, \cdots, N$}
	\STATE {\bfseries data size design}: update $D_k(t)$ by solving Problem \ref{relaxed-problem} (e.g., using gradient ascent).
        \STATE {\bfseries reward design}: update $R_k(t)$ via (\ref{reward-amount})-(\ref{reward-condition}).
	\ENDFOR
	\STATE update iteration index: $t\leftarrow t+1$.
	\UNTIL{$\boldsymbol{R}(t)$ and $\boldsymbol{D}(t)$ converge.}
 }
	\end{algorithmic}  
\end{algorithm} 
Analyzing Algorithm 1's convergence is challenging. First, Problem \ref{server-mechanism-problem} is a mixed-integer program with a large search space.
Second, $\tilde{F}$ is not jointly concave in reward $\boldsymbol{R}$ and data size $\boldsymbol{D}$. Nevertheless, with another mild assumption, we can analyze the algorithm convergence and complexity.\footnote{The optimality analysis is an open problem and left to future work, as the mechanism design is a challenging non-concave and mixed-integer program.}
\begin{Assump}\label{biconcave-assumption}
$\tilde{F}$ is a bi-concave function in $N^P$ and $\boldsymbol{D}$, and satisfies the KL property.
\end{Assump}
Assumption \ref{biconcave-assumption} means that the ensemble accuracy concavely increases in the number of participating learners and the data size. Our experiments in Fig. \ref{tildeF} are consistent with this assumption.  The KL property implies the function is relatively steep around the critical point, and is  satisfied by a wide class of non-convex (and even non-smooth) functions \cite{li2019alternating}. 

\begin{The}\label{convergence}
Under Assumptions \ref{monotone-assumption}-\ref{biconcave-assumption}, Algorithm 1 converges. 
\end{The}
Theorem \ref{convergence} is proven by first transforming the decisions of $\boldsymbol{R}$ to the number of participating learners $N^P$. Then, the result of the proof follows that of Theorem 2.9 in \cite{xu2013block}.
Our numerical experiments in Fig. \ref{algorithm-convergence} also show that the algorithm converges under various parameters. 

\begin{The}\label{complexity}
Algorithm 1 has a complexity $\mathcal{O}(N\log N + L N D^{\rm max})$, where $L$ is the number of alternating iterations. 
\end{The}
Theorem \ref{complexity} is proven by showing that sorting learners takes $\mathcal{O}(N\log N)$, and solving reward and data size (e.g., via gradient ascent) in each iteration takes $\mathcal{O}(D^{\rm max})$.

Theorem \ref{complexity} shows that Algorithm 1 is polynomial in both the number of learners and the maximum data size. This implies that our algorithm is scalable and can be used in practice with a large number of learners and a large dataset. 

\section{Experimental Results}\label{numerical}
We conduct numerical experiments to validate our analysis and draw new insights. In Section \ref{DF-experiments}, we study the property of the surroagte function $\tilde{F}$ (see (\ref{transformed-double-fault})). In Section \ref{algorithm-performance}, we study the convergence of Algorithm 1. In Section \ref{impact-valuation}, we study the impact of the server's valuation on the mechanism performance.

Our experiments are based on the MNIST dataset \cite{mnist}. The dataset contains 70k images of handwritten digits in which 60k are training data and 10k are test data. Our codes are made public in \cite{code}.

\subsection{Property of Surroagte Function $\tilde{F}$}\label{DF-experiments}
\begin{figure}[t]
  \vspace{-6mm}
	\centering
	\subfloat[Surrogate ensemble accuracy $\tilde{F}$.]{\includegraphics[width=0.48\linewidth]{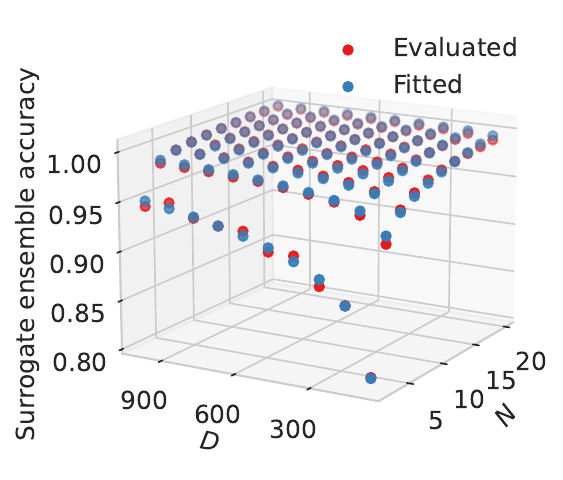}
		\label{tildeF}}
	\hfil
	\subfloat[True ensemble accuracy.]{\includegraphics[width=0.48\linewidth]{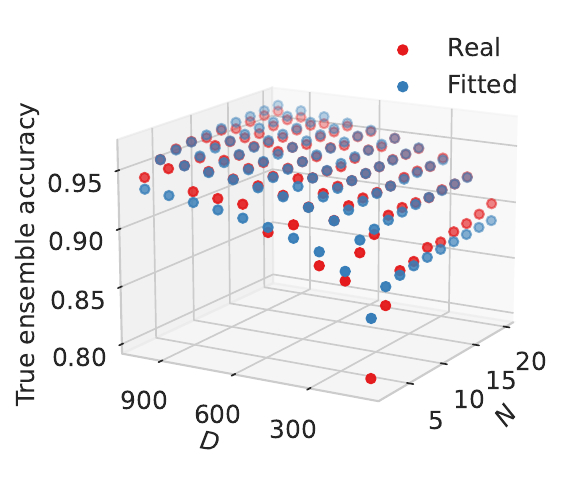}
		\label{ensemble-accuracy}}
	\caption{Impact of learner number and data size on the surrogate and true ensemble accuracy.}
	\label{DF-ensemble}
\end{figure}

We numerically investigate the properties of $\tilde{F}$ and show that it is a good surrogate to the true ensemble accuracy. Here, the true ensemble accuracy is calculated using the aggregated predictions from all  learners' model output via majority voting. In the experiments, we use $N=100$ and assign each learner a dataset with size $D_i\in [200,1000]$ using sampling with replacement. We plot $\tilde{F}$ and the true ensemble accuracy in Fig. \ref{DF-ensemble}. We also use curve fitting to simulate both $\tilde{F}$ and the true ensemble accuracy, and the function takes the form: $(a\log(Nb+c)+d)(e\log(\frac{f}{N}\sum_{i \in \mathcal{N}}D_i+g) + h)$.\footnote{The detailed values of $\{a,b,c,d,e,f,g,h\}$ for both $\tilde{F}$ and true ensemble accuracy are given in the online technical report \cite{appendix}.}

In Fig. \ref{DF-ensemble}, we observe that as learners use more data, the improvements of both the surrogate and true ensemble accuracy are marginally decreasing. 
Also, as more learners participate in DEL, the ensemble accuracy concavely increases.

To further evaluate whether $\tilde{F}$ is a good surrogate to the true ensemble accuracy, we calculate the widely adopted Pearson coefficient \cite{pancholi2022source} between the two functions. The Pearson coefficient takes values in $[-1,1]$, where values close to 1 (-1, respectively) indicate strong positive (negative, respectively) correlations, and values close to $0$ indicate weak correlations. The Pearson coefficient in our experiment is $0.685$, which implies a strong positive correlation. 

We summarize the key observations as follows:
\begin{Obs}
(i) Both $\tilde{F}$ and true ensemble accuracy concavely increases in the learner number and the data size. \\
(ii) The surrogate $\tilde{F}$ has a strong positive correlation with the true ensemble accuracy. 
\end{Obs}

\subsection{Algorithm Convergence}\label{algorithm-performance}
In this subsection, we study the convergence of the proposed algorithm.\footnote{Here we do not study the optimality property as the search of global optimum is experimentally infeasible given a huge search space, i.e.,  $2^N \cdot (D^{\rm max})^N$, where $D^{\rm max}=6\cdot 10^4$. We leave the algorithm development to find the global optimum to future work.} 
In the experiments, we initialize 100 base learners and set $\alpha_i+\beta_i$ for each learner $i$ uniformly in [1e-5, 1e-3], and initialize $\boldsymbol{R}=\boldsymbol{0}$ and $\boldsymbol{D}=\boldsymbol{500}$. 
  \begin{figure}[t]
	\centering
	\subfloat[Optimal number of participating learners]{\includegraphics[width=0.31\linewidth]{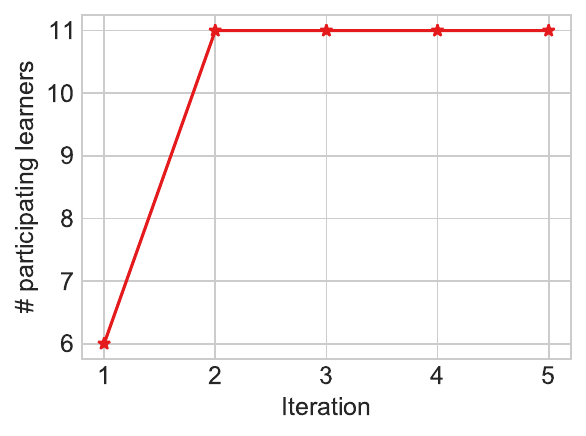}
		\label{N-convergence}}
	\hfil
	\subfloat[Optimal data size]{\includegraphics[width=0.31\linewidth]{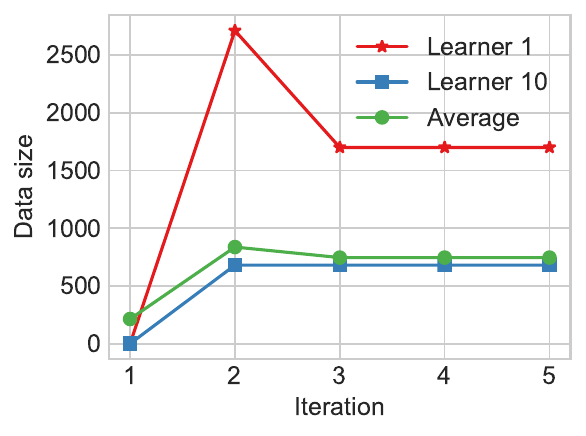}
		\label{D-converegnce}}
        \subfloat[Number of iterations needed for convergence]{\includegraphics[width=0.31\linewidth]{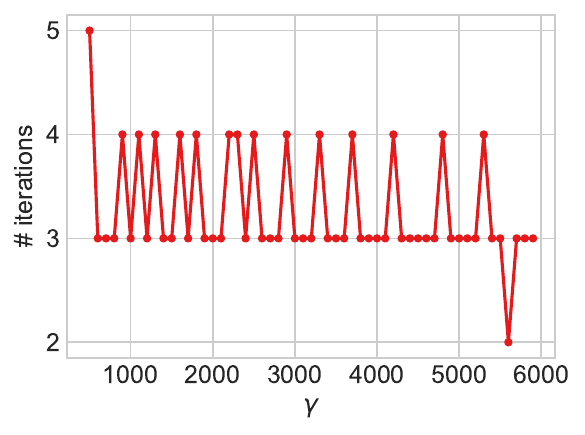}
		\label{iteration-convergence}}
	\caption{Algorithm convergence.}
	\label{algorithm-convergence}
 \vspace{-3mm}
\end{figure}
We plot how the optimal number of learners and data size change with the iteration index $t$ in Fig. \ref{N-convergence} and Fig. \ref{D-converegnce}, respectively. 
We further test the convergence under different values of $\gamma$ and plot the number of iterations needed for convergence in Fig. \ref{iteration-convergence}. The results show that our algorithm achieves fast convergence within less than 5 iterations on average.

  \begin{figure}[t]
	\centering
        \subfloat[Ensemble accuracy]{\includegraphics[width=0.31\linewidth]{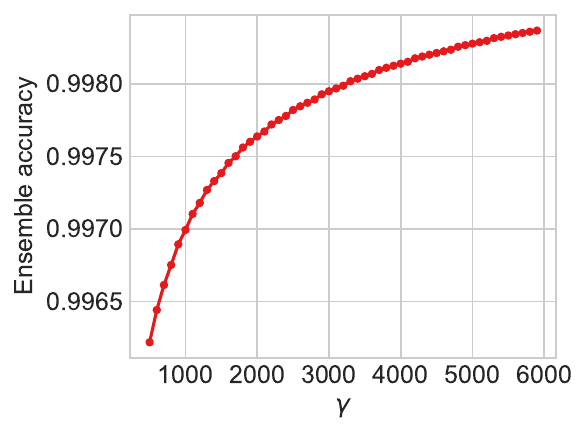}
		\label{accuracy-valuation}}
	\subfloat[Optimal number of participating learners]{\includegraphics[width=0.31\linewidth]{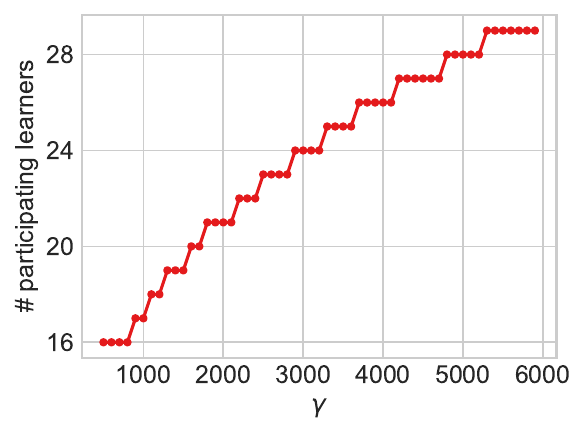}
		\label{N-valuation}}
	\hfil
        \subfloat[Learner diversity]{\includegraphics[width=0.31\linewidth]{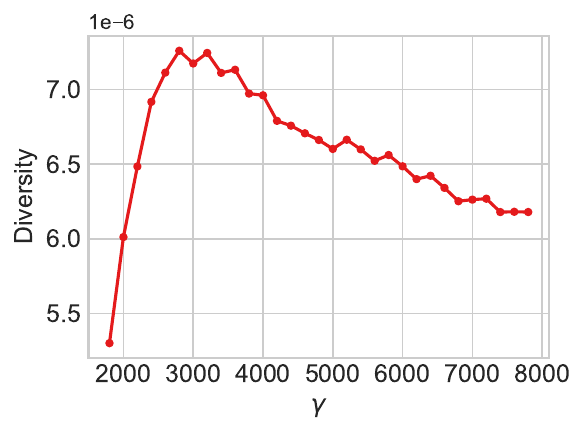}
		\label{diversity-valuation}}
	\caption{Impact of server valuation on mechanism performance. }
	\label{impact-valuation-f}
\end{figure}


\subsection{Impact of Server's Valuation on Mechanism Performance}\label{impact-valuation}
In this subsection, we study how the server's valuation on the ensemble accuracy affects the mechanism performance. In the experiment, we consider $\alpha_i+\beta_i$ for each learner $i$ uniformly distributed in [1e-5, 1e-3] and change $\gamma \in [500,8000]$. Fig. \ref{impact-valuation-f} plots how the true ensemble accuracy, the optimal number of participating learners, and diversity (the first term in (\ref{transformed-double-fault})) depend on the server's valuation $\gamma$. 

In Fig. \ref{accuracy-valuation}, we observe that as $\gamma$ increases, the resulting ensemble accuracy (after mechanism optimization) increases. The server will incentivize more learners (see Fig. \ref{N-valuation}) to participate in DEL, leading to a higher ensemble accuracy. 

Counter-intuitively, we observe in Fig. \ref{diversity-valuation} that the trend of learners' diversity  first increases and then decreases in server's valuation $\gamma$. When $\gamma$ is small (e.g., $\gamma=2000$), the server incentivizes only a few learners. To achieve a high ensemble accuracy, the few learners should not be too diverse, because otherwise their wrong predictions cannot be corrected by the few remaining learners. When $\gamma$ increases (e.g., $\gamma=3000$), the server incentivizes a larger learner pool which is more robust to wrong predictions. The server is better off diversifying the learners so that they can learn from different mistakes, leading to a higher ensemble accuracy. As $\gamma$ keeps growing (e.g., $\gamma=5000$), the server incentivizes even more learners, but their diversity value slightly decreases. 
 This is because it is difficult to reach a prediction consensus when a large number of learners are too diverse. As a result, one needs to ensure a moderate level of diversity to achieve the best ensemble accuracy.

We summarize the above observations below.
\begin{Obs}
(i) The ensemble accuracy and the optimal number of participating learners increase in $\gamma$.(ii) When the number of participating learners is large, the server prefers a lower level of learner diversity to achieve a higher accuracy. 
\end{Obs}

\section{Conclusion}\label{sec: conclusion}
This paper presents the first study on the incentive mechanism design for distributed ensemble learning. The mechanism design is a challenging mixed-integer program with a large search space. To address this issue, we propose an alternating algorithm that iteratively updates the data size and reward for heterogeneous learners. We prove that the algorithm converges and is scalable to large distributed systems. Numerical experiments using MNIST dataset show an important insight: when the number of participating learners is large, the server prefers a lower level of learner diversity to achieve a higher ensemble accuracy.

There are a few exciting directions for future work. For example, it would be interesting to extend the mechanism to the incomplete information case where the server does not know each learner's cost information. One can resort to Bayesian game-theoretical tools or auction mechanisms. Another interesting direction is to study the mechanism design for other ensemble learning frameworks such as boosting and stacking. 

\bibliographystyle{IEEEtran}
\bibliography{GLOBECOM-Final}

\end{document}